\documentclass{aa}

\usepackage{graphicx}
\usepackage{txfonts}

\newcommand{\Mpc}{$h^{-1}$\thinspace Mpc}

\newcommand{\etal}{{\rm et al.~}}

\def\apjl{ApJL}
\def\apjs{ApJS}

\def\aap{A\&A}

\begin{document}   

\title{Superclusters of galaxies from the 2dF redshift survey.  \\
II. Comparison with simulations} 

\author{ J. Einasto\inst{1} \and M. Einasto\inst{1} \and
 E. Saar\inst{1} \and  E. Tago\inst{1} \and L. J. Liivam\"agi\inst{1} \and
 M. J\~oeveer\inst{1} \and I. Suhhonenko\inst{1} \and
 G. H\"utsi\inst{1} \and J. Jaaniste\inst{2} \and P.
 Hein\"am\"aki\inst{3} \and V. M\"uller\inst{4} \and A. Knebe\inst{4}
 \and D. Tucker\inst{5}}

\institute{Tartu Observatory, EE-61602 T\~oravere, Estonia
\and 
Estonian University of Life Sciences
\and 
Tuorla Observatory, V\"ais\"al\"antie 20, Piikki\"o, Finland 
\and
Astrophysical Institute Potsdam, An der Sternwarte 16,
D-14482 Potsdam, Germany
\and
 Fermi National Accelerator Laboratory, MS 127, PO Box 500, Batavia,
IL 60510, USA
}

\date{ Received 2006; accepted} 

\authorrunning{J. Einasto et al.}

\titlerunning{2dFGRS superclusters}

\offprints{J. Einasto }

\abstract {We investigate properties of superclusters of galaxies found
  on the basis of the 2dF Galaxy Redshift Survey, and compare them
  with properties of superclusters from the Millennium Simulation.
  We study the dependence of various characteristics of superclusters
  on their distance from the observer, on their total luminosity, and
  on their multiplicity.  The multiplicity is defined by the number of Density
  Field (DF) clusters in superclusters.  Using the multiplicity we
  divide superclusters into four richness classes: poor, medium, rich
  and extremely rich.  We show that superclusters are asymmetrical and
  have multi-branching filamentary structure, with the degree of asymmetry
  and filamentarity being higher for the more luminous and richer
  superclusters. The comparison of real superclusters with 
  Millennium superclusters shows that most properties of simulated
  superclusters agree very well with real data, the main differences being 
  in the luminosity and multiplicity distributions.

\keywords{cosmology: large-scale structure of the Universe -- clusters
of galaxies; cosmology: large-scale structure of the Universe --
Galaxies; clusters: general}

}

\maketitle

\section{Introduction}

The largest non-percolating galaxy systems are superclusters of galaxies which
contain clusters and groups of galaxies along with their surrounding galaxy
filaments.  Superclusters evolve slowly and contain information about the very
early universe; thus their properties can be used as cosmological probes to
discriminate between different cosmological models.  Early studies of
superclusters were based on galaxies and groups (see the reviews by Oort
\cite{oort83} and Bahcall \cite{bahcall88}).  All-sky catalogues of
superclusters were complied by Zucca \etal (\cite{z93}), Kalinkov \& Kuneva
(\cite{kk95}), Einasto \etal (\cite{e1994}, \cite{e1997}, \cite{e2001},
hereafter E01) using galaxy cluster catalogues by Abell (\cite{abell}) and
Abell \etal (\cite{aco}).

In fact, superclusters consist of galaxy systems of different richness
classes -- from single galaxies, galaxy groups and filaments to rich
clusters of galaxies.  This has been realised long ago (J\~oeveer,
Einasto \& Tago \cite{jet78}, Gregory \& Thompson \cite{gt78}) and has
been confirmed by recent studies of superclusters using deep new 
galaxy surveys, such as the Las Campanas Galaxy Redshift Survey, the 2
degree Field Galaxy Redshift Survey (2dFGRS, Colless et
al. \cite{col01}, \cite{col03}), and the Sloan Digital Sky Survey
(SDSS) (Doroshkevich et al.  \cite{dor01}, Einasto et al. \cite{e03a},
\cite{e03b}; Erdogdu et 
al. \cite{erd04}; Porter and Raychaudhury \cite{pr05}; and Einasto
et al. \cite{e06}, hereafter Paper I).  These new surveys have shown
that morphological properties of galaxies depend on the large-scale
cosmological environment: Einasto et al. (\cite{einm03c}, \cite{e03a},
\cite{ein05b}, hereafter E05b); Balogh et al. (\cite{balogh04}); Croton et
al. (\cite{croton05a}); Lahav (\cite{lahav04} and \cite{lahav05});
Ragone et al. (\cite{ragone04}).  The densest global environment is
provided in superclusters of galaxies.

\begin{figure*}[ht]
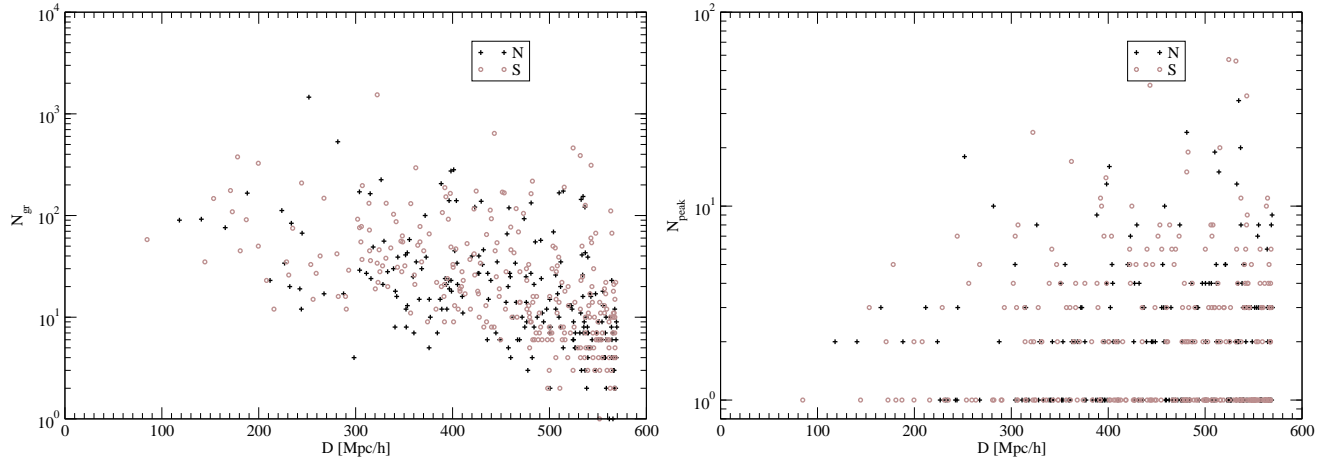

\centering
\resizebox{.48\textwidth}{!}{\includegraphics*{scl_NSgr8_Ngr-dist.eps}}
\resizebox{.48\textwidth}{!}{\includegraphics*{scl_NSgr8_Npeak-dist.eps}}
\\
\caption{The multiplicity of superclusters at various distances from
  the observer. Here we have used groups of galaxies in the left panel
  and DF-clusters in the right panel as supercluster multiplicity
  indicators. }
\label{fig:1}
\end{figure*}

The goal of this paper is to investigate properties of real superclusters and
to compare them with simulated superclusters found in numerical simulations of
the evolution of the structure of the Universe.  We shall use the supercluster
catalogue of Paper I and catalogues of simulated superclusters based on the
Millennium Simulation of Springel et al. (\cite{springel05}) (see also Gao et
al. \cite{gao05} and Croton et al. \cite{croton05b}).  Traditionally in
supercluster studies the multiplicity and the shape is discussed and used as a
cosmological probe (Basilakos et al. \cite{bpr01}, Kolokotronis et al.
\cite{kbp02}, Basilakos \cite{bas03}, Wray et al. \cite{wray06}).  Additional
parameters of galaxy systems in high- and low-density environment
(supercluster and field populations) were investigated by Doroshkevich et al.
(\cite{dor01}) using the minimal spanning tree and inertia tensor methods. In
this paper we shall discuss also a number of supercluster properties, such as
the luminosity function; the maximal, minimal and effective diameters and
their ratios; the geometric and dynamical centres and their differences; and
the main clusters and galaxies.    We define the multiplicity of superclusters
by the number of 
Density Field (DF) clusters, and divide superclusters according to their
multiplicity into four richness classes: poor, medium, rich and extremely
rich.  Properties of galaxies in superclusters, including properties of main
galaxies and main clusters, shall be investigated in a separate paper by
Einasto et al.  (\cite{e2006c}, Paper III).

The paper is composed as follows. Section 2 describes the data used for
the analysis. In Section 3 we  discuss multiplicities of
superclusters and define supercluster richness classes.  In Section 4
we  study properties of superclusters and compare these
properties with those of simulated superclusters.  We consider
the dependence of supercluster properties on the distance from the
observer, on the total luminosity,  and on the richness.  In Section 5 we
discuss our results and compare our lists of superclusters with
superclusters found in earlier studies.  In the last sections we give
our conclusions. The catalogue of superclusters is available
electronically at the web-site
\texttt{http://www.aai.ee/$\sim$maret/2dfscl.html}.

\section{Data}

We have used in this analysis supercluster catalogues published by
Einasto et al. (Paper I).  The main data used in this paper can be found in
the electronic supplement of this paper; some unpublished data on real
and model supercluster samples are also used.  The main data are the
following. 

The geometric data are: the supercluster distance $d$, and the minimal,
maximal, and effective diameters of the supercluster, $D_{min}$,
$D_{max}$, $D_{e}$. The minimal diameter is the shortest size of the
supercluster along rectangular coordinates $x,~y,~z$, calculated in
rectangular equatorial coordinates (for details see Paper I), the maximal
diameter is the diagonal of the box containing the supercluster along
rectangular coordinates, and the effective diameter is the diameter of
the sphere whose volume equals that of the supercluster. We further use the
ratio of the mean to effective diameter $\epsilon_0 = D_m/D_e$, where
$D_m = D_{max}/3^{1/3}$ is the mean diameter.  This parameter
characterizes the compactness of the system. A further geometrical
parameter is the center offset, which is the difference
between the geometric center (as defined by the $x,~y,~z$-coordinates
of the density field above the threshold density), and the dynamic
centre, identified with the center of the main (most luminous) cluster
of the supercluster. This parameter characterises the asymmetry of the
supercluster.

The physical data are: the peak and mean density of the supercluster (in units
of the mean luminosity density); the number of galaxies and groups (from the
catalogue by Tago \etal \cite{tago06}, hereafter T06); the multiplicity of the
supercluster, defined as the number of DF-clusters in it (for definition of
DF-clusters see below Sect. 3); and the total luminosity of the supercluster,
$L_{tot}$, expressed in Solar units in the ${\rm b_j}$ filter passband. The
total luminosity of a supercluster is calculated by summing the estimated
total luminosities of its member galaxies and groups.

We use corresponding characteristics for simulated superclusters using the
Millennium Simulation by Springel et al. (\cite{springel05}).

\section{Multiplicities of superclusters and supercluster richness
 classes}

{\begin{table*}[ht]
\caption{Data on 2dFGRS and Millennium Simulation superclusters of
  various richness }
\begin{tabular}{lrrrrrrrrr} 
\hline 
\\ 
Name  &    &      &          &    2dF  &         &     &          &  Mill     &             \\ 
\hline 
\\ 
      &  Rich   &  N   &   1st    &  med    & 4nd     &     N   &    1st   & med   &  4nd        \\
\\ 
\hline 
\\
$L_{tot}$ &  P  &   365 &  5.157e11& 8.302e11& 1.347e12&    1432 &  1.554e12&  2.425e12& 3.962e12 \\
          &  M  &   151 &  1.526e12& 2.583e12& 4.267e12&     283 &  6.028e12&  1.028e13& 1.720e13 \\
          &  R  &    18 &  1.055e13& 1.528e13& 1.866e13&      18 &  2.695e13&  3.852e13& 4.278e13 \\
          &  E  &     9 &  2.024e13& 4.320e13& 4.975e13&       1 &          &  1.048e14&          \\
                                                                        &          &          \\
$M_m$     &  P  &       &   -20.66 & -20.94 & -21.30 &         &   -22.14 &  -22.44  &-22.77    \\
          &  M  &       &   -20.73 & -21.02 & -21.34 &         &   -22.61 &  -22.87  &-23.13    \\
          &  R  &       &   -21.16 & -21.45 & -21.74 &         &   -23.00 &  -23.20  &-23.42    \\
          &  E  &       &   -21.09 & -21.31 & -21.57 &         &          &  -23.71  &          \\
                                                                        &          &          \\
$\delta_m$&  P  &       &    5.07  &  5.37   &  5.99   &         &  5.03    &   5.36   &  5.95    \\
          &  M  &       &    5.42  &  5.89   &  6.46   &         &  5.57    &    6.14  &    6.88    \\
          &  R  &       &    6.25  &  6.90   &  7.19   &         &  6.45    &   6.71   &  7.35    \\
          &  E  &       &    6.91  &  7.26   &  7.87   &         &          &   7.35   &          \\
                                                                        &          &          \\
$\delta_p$&  P  &       &    5.140 &  5.780 &   7.330    &         &   5.200 &  5.930  &  7.533  \\
          &  M  &       &    5.675 &  6.740 &   9.070    &         &   5.880 &  8.190  & 11.840    \\
          &  R  &       &    8.643 & 10.400 &  12.460    &         &   10.26 &  11.72  &  16.81  \\
          &  E  &       &    11.07 &  12.29 &   16.33    &         &   11.33 &  11.33  &  11.33  \\
\\
$D_{max}$ &  P  &       &   13.93  & 17.49   & 21.95   &         & 14.07    &   18.22  &  23.69   \\
          &  M  &       &   27.81  & 34.07   & 42.61   &         & 31.49    &    41.68 &    53.00   \\
          &  R  &       &   75.61  & 84.56   & 90.48   &         & 73.80    &   85.32  & 106.30   \\
          &  E  &       &   93.11  &126.40   &153.90   &         &          &  136.4   &          \\
                                                                        &          &          \\
 $D_e$    &  P  &       &    7.51  &  9.58   & 11.70   &         &  7.71    &    9.87  &  12.20   \\
          &  M  &       &   12.21  & 14.82   & 17.70   &         & 13.80    &    16.66 &    19.88   \\
          &  R  &       &   24.90  & 27.10   & 28.49   &         & 23.30    &   25.96  &  27.24   \\
          &  E  &       &   29.73  & 39.44   & 40.90   &         &          &   36.68  &          \\
                                                                        &          &          \\
$\Delta_o$&  P  &      &    2.39  & 3.41    &  4.33   &         & 1.57     &     2.55  &    4.06   \\
          &  M  &      &    5.05  & 6.59    &  8.70   &         & 5.42     &     8.27  &   12.50   \\
          &  R  &      &    13.2  & 15.2    &  23.5   &         & 13.6     &   23.49  &  27.89   \\
          &  E  &      &    15.5  & 19.2    &  23.4   &         &          &   26.26  &          \\
                                                                        &          &          \\
$\epsilon_0$ &  P &    &    1.02  &  1.06   &  1.12   &         & 1.02     &   1.06   & 1.15     \\
          &  M  &      &    1.26  &  1.33   &  1.49   &         & 1.32     &    1.44  &   1.58     \\
          &  R  &      &    1.65  &  1.77   &  1.83   &         & 1.79     &   1.96   & 2.13     \\
          &  E  &      &    1.81  &  1.95   &  2.22   &         &          &   2.15   &          \\

\\
\hline
\label{tab:SCL-dat}
\end{tabular}
\end{table*}
}

Peaks of the luminosity density have been calculated using luminosities of
groups corrected for the effect of luminosity bias. These data are used to
find DF-clusters -- peaks of the density field smoothed with an Epanechnikov
kernel of radius 8~\Mpc.  The use of the Epanechnikov kernel is preferable
since it has no low-density wings as the Gaussian kernel. The number density
of DF-clusters is 57 and 58 per million (\Mpc)$^3$, respectively, in the
2dFGRS Northern and Southern regions, using a threshold density of 5.0 in
units of the mean density.  For the Millennium Simulation density field
DF-clusters have been found for a density threshold of 5.5; for this limit the
number density of DF-clusters is 25 per million (\Mpc)$^3$. For comparison we
note that the number density of Abell clusters is 26 per million (\Mpc)$^3$
(Einasto et al. \cite{e1997}).  If a higher threshold density is used, then
the number density of DF-clusters in 2dFGRS samples becomes very close to the
Millennium Simulation and Abell cluster density estimates.  In other words,
DF-clusters can be considered as an equivalent to Abell clusters.  We searched
the neighbourhood of DF-clusters for galaxies and groups of galaxies.  This
search shows that all DF-clusters lie in a high-density environment: in a box
of radius 3~\Mpc\ around a DF-cluster there are, depending on the distance
from the observer, up to 60 2dFGRS groups and field galaxies outside groups.

The multiplicity of Abell superclusters was derived by Einasto et
al. (\cite{e1994}).  The distribution of multiplicities is rather
close to the distribution found in the present paper. 
The multiplicity of simulated superclusters was studied  by Wray
et al. using rich DM-halos.  The multiplicity function was derived for
a number of linking lengths.  The linking length $L =
5$~\Mpc\ corresponds approximately to our accepted threshold
density. The integrated multiplicity function is presented in their
Fig. 3.  The case $L=5$~\Mpc\ is rather similar to the multiplicity
function of our DF-clusters.  We shall discuss the multiplicity
function and its cosmological consequences in  separate papers by Saar
et al (\cite{see06}) and Einasto et al. (\cite{esh06}).

\begin{figure*}[ht]
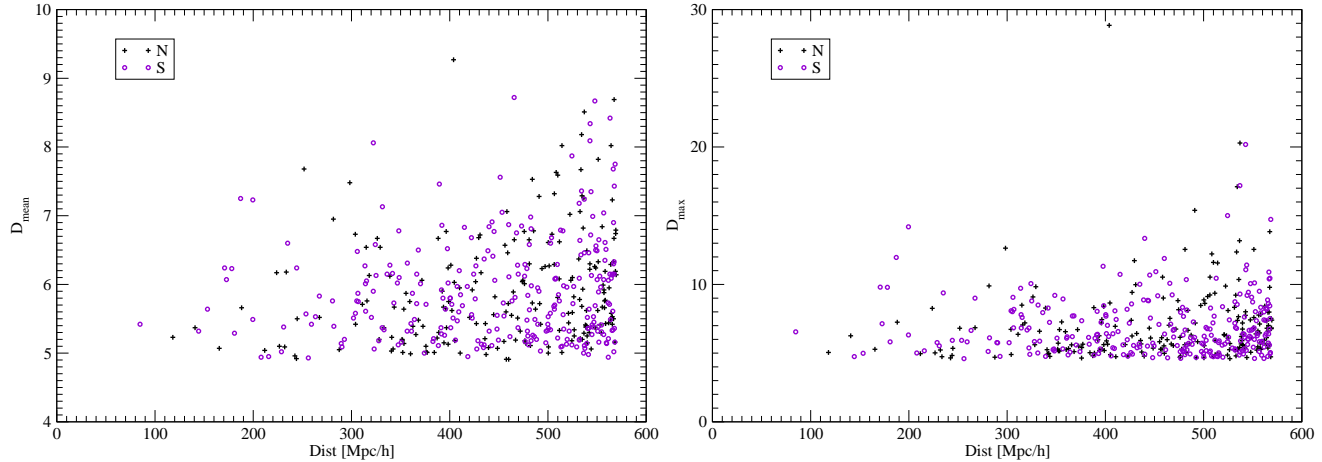

\centering
\resizebox{.48\textwidth}{!}{\includegraphics*{scl_NSgr8_Dmean-dist.eps}}
\resizebox{.48\textwidth}{!}{\includegraphics*{scl_NSgr8_Dmax-dist.eps}}
\\
\caption{The mean  (left panel) and the maximal density (right
  panel) of superclusters at various distances from the observer.  
}
\label{fig:2}
\end{figure*}

A natural division of superclusters into poor, medium and rich systems
would be according to the total luminosity.  However, it is not always easy to
determine the total luminosity of superclusters.  Another  
 possibility to characterize the richness of superclusters is the
number of high-density peaks, in our case DF-clusters.  As we have seen above,
the number of galaxies and groups is not suited for this purpose since
both these numbers are influenced by selection effects.
The number of DF-clusters is
proportional to the supercluster total luminosity: the relationship is
linear in the log-log diagram. Multiplicity of superclusters is easily
determined from observation, also for superclusters found using Abell
clusters or DM-halos in simulations.

We shall use the supercluster multiplicity to divide superclusters
into four richness classes.  We call superclusters which have less
than 3 DF-clusters poor, superclusters with 3 to 9 DF-clusters
medium rich, superclusters with 10 to 19 DF-clusters rich, and with
20 or more DF-clusters extremely rich superclusters. As shown
above, DF-clusters are of the Abell class.  As a prototype of a poor
supercluster we can use the Virgo or Coma Supercluster, containing
one and two Abell class clusters, respectively.  A characteristic
supercluster of medium richness class is the Perseus-Pisces
Supercluster; its visible part contains 3 Abell clusters, but part of
the supercluster is hidden in the zone of avoidance.  Examples of
nearby rich superclusters are the Leo-Sextans and
Horologium-Reticulum Superclusters, and also the supercluster SCL126
(see Fig.~\ref{fig:scl} below).  To the class of extremely rich
superclusters belongs the Shapley Supercluster.

We have used these richness classes and determined for all richness classes
median values of various parameters, determined in Paper I for 2dFGRS
superclusters.  To get an idea of the spread in these parameters we calculated
also the 1st and 4nd quantile of the distribution.  Data are given in
Table~\ref{tab:SCL-dat}, both for the 2dFGRS and Millennium Simulation (sample
Mill.F8) superclusters.  Here $N$ is the number of superclusters of the
respective richness class, denoted by P, M, R and E for poor, medium, rich,
and extremely rich superclusters, respectively (the numbers are given only
once, they are identical for all quantities given in the Table).  The
quantities are: the luminosity $L_{tot}$ (in Solar units); the absolute
magnitude of the main galaxy $M_m$ (in the {\rm $b_j$} filter passband for
2dFGRS, and in the {\rm g} filter passband for the Millennium Simulation); the
mean and maximal densities, $\delta_m$ and $\delta_p$, respectively (in units
of the mean density); the maximal diameter $D_{max}$ (diagonal of the
rectangular box around the supercluster) and the effective diameter $D_e$ (the
diameter of the sphere equal in volume to that of the supercluster, both in
\Mpc); the center offset parameter $\Delta_o$ (the distance between the
geometric and dynamic center, also in \Mpc); and the ratio of the effective to
the mean diameter $\epsilon_0$.

The number of DF-clusters in superclusters is on average the same in nearby
and distant superclusters (see Fig.~\ref{fig:1}).  There is a strong
dependence of the number of DF-clusters with the total luminosity of
superclusters.  A look at Table~\ref{tab:SCL-dat} shows that about 2/3 of the
superclusters are poor, both in the real and simulated samples -- i.e. they
belong to the same supercluster richness class as the Local and Coma
Superclusters.  With increasing total luminosity the number of DF-clusters --
or the multiplicity of superclusters -- increases.

However, there exists a remarkable difference between the distribution of
supercluster richness classes for the real and simulated samples.  For the
2dFGRS supercluster sample, 5\% of the superclusters are of richness class R
and E, and the richest superclusters have up to about 60 DF-clusters.  In
contrast, only 1\% of the Millennium Simulation supercluster sample has this
richness class, and the richest simulated superclusters have only 20
DF-clusters. This difference in the richness of real and simulated
superclusters is one of the major results of our study.  A further discussion
of this problem is given by Saar et al.~(\cite{see06}).

\begin{figure*}[ht]
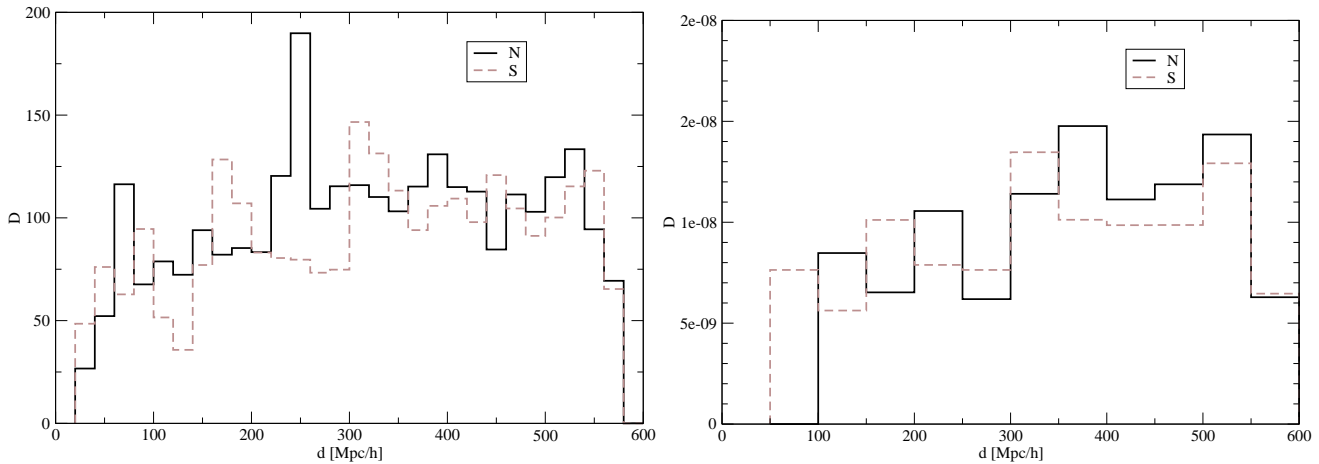

\centering
\resizebox{.48\textwidth}{!}{\includegraphics*{2dfNS7_Dens-dist.eps}}
\resizebox{.48\textwidth}{!}{\includegraphics*{scl_2dfNSgr_Dens-dist.eps}}
\\
\caption{Left panel: the mean luminosity density in the Northern and Southern
  regions of the 2dF as a function of distance.  Right panel: the mean number
  density of 2dF superclusters  as functions of distance. 
}
\label{fig:3}
\end{figure*}

\section{Properties of real and simulated superclusters}

\subsection{Supercluster properties as a function of distance}

The major problem in using flux-limited galaxy samples is the
magnitude selection effect.  Faint galaxies and groups are not seen at
large distances, thus it is expected that the number of visible
galaxies and groups in superclusters decreases with distance.  This is
what is actually observed: there exists a lower limit of the number of galaxies
and groups in superclusters, and this limit decreases with distance
exponentially (Fig.~\ref{fig:1}).  This selection effect distorts
the number of galaxies and groups as supercluster richness indicators.
For this reason we have used  as richness indicator the number of local
high-density peaks -- DF-clusters (see right panel of
Fig.~\ref{fig:1}).  DF-clusters have been found using luminosities of
galaxies and groups corrected for the effect of luminosity bias. (See
Paper I for details of the reconstruction of estimated true total
luminosities of DF-clusters and superclusters.)  The right panel of
Fig.~\ref{fig:1} shows that the number of DF-clusters in superclusters
-- the supercluster multiplicity -- is in the mean the same in nearby
and distant superclusters. Due to the volume-effect there are no very rich
superclusters at small distance $D < 300$~\Mpc.

In Fig.~\ref{fig:2} we show the mean and maximal luminosity density of
superclusters at various distances from the observer, for the Northern and
Southern regions. We see that the distributions are rather uniform --
both densities are distributed similarly irrespective of distance.

The left panel of Fig.~\ref{fig:3} shows the mean luminosity density of the
whole 2dFGRS sample, separately for the Northern and Southern regions.  There
are several peaks due to very rich superclusters, but in general the
luminosity density is approximately constant. The mean number density of
superclusters is shown in the right panel of Fig.~\ref{fig:3}; it is also
approximately the same at various distances from the observer.  2dFGRS samples
are wedges and are rather thin at small distance from the observer. In such
thin regions the identification of superclusters is problematic and hence the
spatial density of superclusters at small distances is rather low.  This is
the only known distance-dependent selection effect in the supercluster sample.

\begin{figure*}[ht]
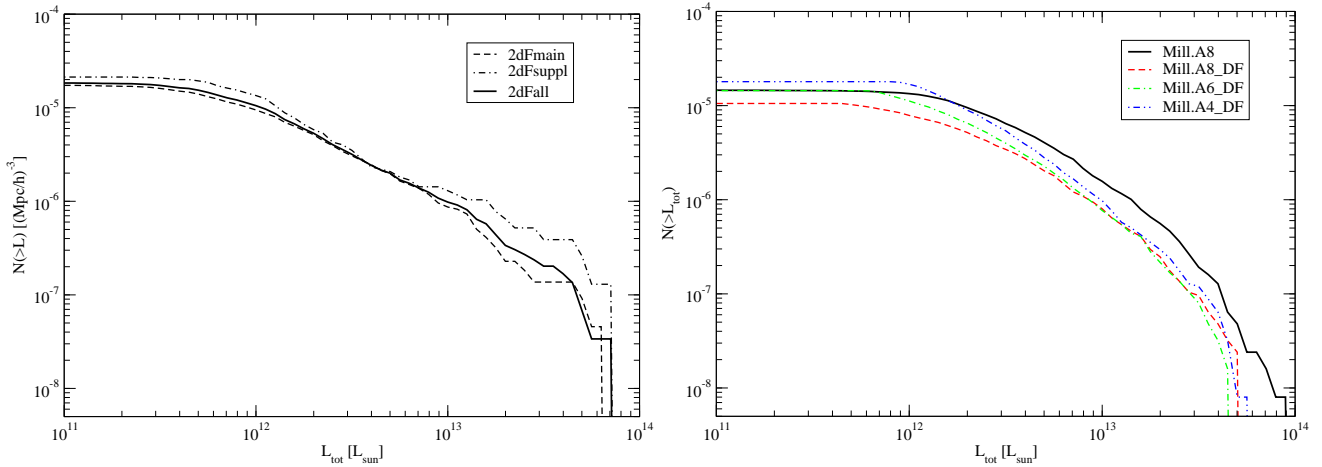

\centering
\resizebox{.48\textwidth}{!}{\includegraphics*{scl_2dfNSgr_all_lumf_b.eps}}
\resizebox{.48\textwidth}{!}{\includegraphics*{scl_M500F8_100_lumf.4.eps}}
\\
\caption{The luminosity functions of superclusters. Left: 2dF
  superclusters, right: Millennium Simulation superclusters. The supercluster
  luminosity function for the 2dFGRS was calculated for the main,
  supplementary, and the total supercluster samples (North and South added),
  plotted with dashed, dot-dashed and solid curve, respectively.  For
  simulated superclusters the solid bold curve is the luminosity function
  calculated by adding luminosities of supercluster member galaxies; dashed,
  dot-dashed and dot-dot-dashed curves correspond to Millennium samples
  Mill.A8, Mill.A6 and Mill.A4, in these cases the luminosity of superclusters
  was found from the density field.}
\label{fig:lumf}
\end{figure*}

\subsection{Luminosities of superclusters}

One of the major goals of the present study was the compilation of a
representative sample of superclusters over a broad range of
luminosities.  This gives us the possibility to derive the luminosity
function of superclusters.  Results of our calculations are shown in
Fig.  \ref{fig:lumf}.  Our dataset is large enough to measure the
luminosity function spanning almost 3 orders in luminosity and spatial
density.  

We have calculated the supercluster luminosity function separately for
the main supercluster sample, which contains superclusters with
distances up to 520~\Mpc, and for the supplementary sample, containing
more distant superclusters.  In both cases Northern and Southern
subsamples were stacked.  The main samples have 379 superclusters, the
supplementary ones 164.  Our analysis has shown that the largest and
most luminous superclusters have extensions beyond the boundaries of
2dFGRS. Thus the total luminosities of most luminous superclusters
(SCL9 and SCL126 from the list by Einasto et al. (\cite{e1997})) are
probably underestimated.
 
Figure~\ref{fig:lumf} shows that there are no large differences between
luminosity functions found for the main and supplementary samples, only the
scatter of the function of the supplementary sample is larger.  This is due to
two factors: the expected total luminosities of groups have random errors due
to large weight factors used in the restoration of the expected total
luminosities, and there are errors in the density field due to inaccuracies in
the expected total luminosities. Due to the second error some superclusters
fall below the threshold density and are missed, and some poor superclusters
are added to our list which actually are fainter than our threshold density
limit. In the mean these errors cancel each other out and increase only the
scatter or supercluster luminosities. There is also a certain systematic
trend: the supplementary list contains, in comparison to the main list, richer
superclusters.

In Paper I we have compared luminosities of simulated superclusters 
calculated from the true density field and from the density field
of simulated flux limited galaxy samples. This comparison has shown 
that luminosities of superclusters found from a flux limited sample are
systematically fainter than 'true' luminosities found from all data.
A detailed analysis is needed to investigate this and other biasing
factors (including a detailed study of the structure of individual
rich superclusters) to have a more accurate estimate of the
supercluster luminosity function.

For comparison we show in the right panel of Fig.~\ref{fig:lumf} the
luminosity function of simulated superclusters extracted from the Millennium
Simulation.  The luminosity function was constructed in two ways.  The solid
line corresponds to the sample Mill.A8 where luminosities of superclusters
were found by adding luminosities of galaxies in superclusters. The
supercluster sample includes systems with a minimal volume limit of 100
Mpc$^3h^{-3}$.  Dashed, dot-dashed and dot-dot-dashed curves were calculated
for models Mill.A8, Mill.A6 and Mill.A4, respectively, by integrating the
luminosity density field inside the contour of the supercluster, multiplied by
the mean luminosity per cell. Supercluster samples Mill.A8, Mill.A6, and
Mill.A4 were found using Epanechnikov kernel with radius 8, 6, and 4~\Mpc,
respectively, see Paper I for details.  In these models the limiting volume
was taken to be 200 Mpc$^3h^{-3}$. We see that there exists a shift in
luminosities between the two sets of data. This shift can be explained by the
fact that, in the luminosity density field, part of the luminosity of galaxies
is smoothed away beyond the threshold contour (see Fig.~\ref{fig:scl}).  For
this reason total luminosities obtained from the density field are
underestimates. To avoid this systematic bias we have used in the calculation
of total luminosities of superclusters estimates based upon luminosities of
supercluster member galaxies and galaxy groups/clusters.

The comparison of luminosity functions of the samples Mill.A8, Mill.A6, and
Mill.A4 shows the effect of using different smoothing lengths in
the calculation of the density field.  The smaller the smoothing
length, the more often outlying parts of superclusters have a tendency to
form independent poor superclusters. This increases the number of poor
superclusters and changes the resulting luminosity function. The
figure shows that these changes are rather modest; in the range of
high luminosities the frequency of superclusters of various 
luminosity is almost constant.  If we use identical smoothing lengths and
selection parameters (threshold density and lower volume limit) for models
and real data, we get comparable supercluster samples.

The mean number-density of superclusters as defined in the present analysis is
well determined: both the Northern and Southern regions of 2dF yield about 17
superclusters per million Mpc$^3 h^{-3}$. A sample with similar selection
parameters for the Millennium Simulation contains 14 superclusters per million
Mpc$^3 h^{-3}$ -- i.e. the number densities of real and simulated
superclusters are very similar.  The number density of rich and extremely rich
superclusters is much lower -- 0.6 and 0.3 per million Mpc$^3 h^{-3}$,
respectively, in the real sample.  If this density is characteristic for the
whole Universe, then we have a chance to observe one extremely rich
supercluster in a volume of 3 million Mpc$^3h^{-3}$. We note that nearby
examples of rich and extremely rich superclusters SCL126 and SCL9 have
actually a higher total luminosity than found in the 2dFGRS, since both are
cut by the observational limits of the survey in declination, as seen from the
density field.  Larger deep galaxy redshift surveys are needed to find more
accurate values of luminosities of these very rich superclusters.

The comparison of luminosity functions of real and simulated
superclusters shows several differences.  First, as noted above, the
number density of rich and extremely rich superclusters in the
simulated sample is much lower, only 0.15 per million Mpc$^3 h^{-3}$.
The other difference between total luminosities of real and simulated
superclusters is in their median luminosity values.  Table
\ref{tab:SCL-dat} shows that median luminosities of simulated
superclusters of all richness classes are higher than median values
of luminosities of 2dF superclusters by a factor of 2 -- 4.  A similar
shift is seen in the luminosity functions of the 2dF and Millennium
Simulation galaxies.  The reason of this discrepancy is not clear.

The third difference between luminosity functions of real and
simulated superclusters is in the shape of the functions: the luminosity
function of 2dFGRS superclusters has an almost constant slope in the $\log
L_{tot}$ vs. $\log N(>L_{tot})$ diagram, whereas the luminosity
function of simulated superclusters has a continuous change of the
slope.

\begin{figure*}[ht]
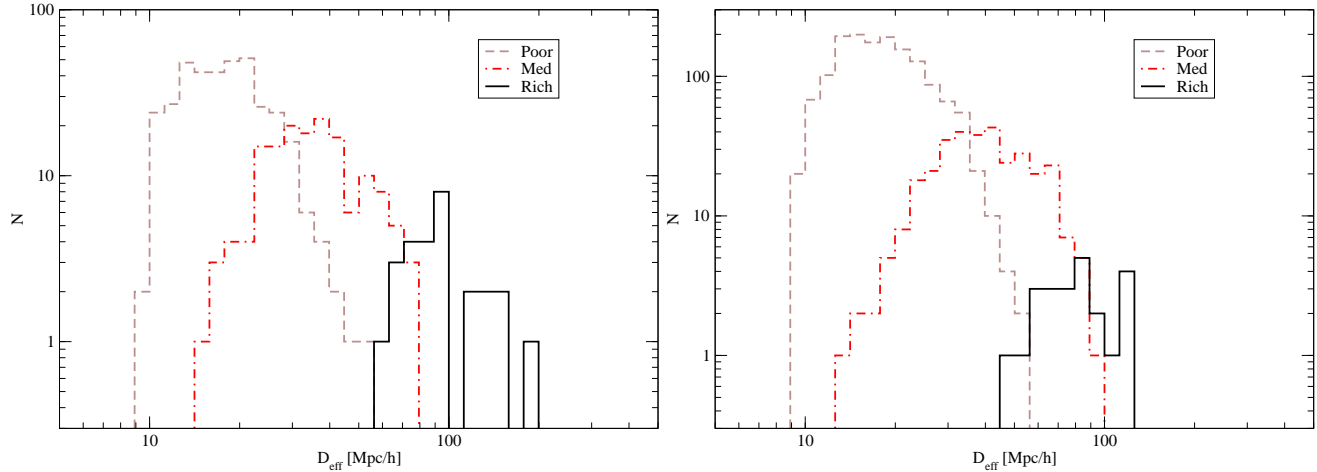

\centering
\resizebox{.48\textwidth}{!}{\includegraphics*{scl_NS_dmax-distr2.eps}}
\resizebox{.48\textwidth}{!}{\includegraphics*{scl_M5_dmax-distr2.eps}}
\\
\caption{The distribution of maximal diameters of poor, medium, and
  rich superclusters, defined by supercluster multiplicity.  Left
  panel shows data for 2dFGRS superclusters (Northern and Southern
  regions together), right panel Mill.A8 superclusters.
}
\label{fig:diam}
\end{figure*}

\begin{figure*}[ht]
\centering
\resizebox{.48\textwidth}{!}{\includegraphics*{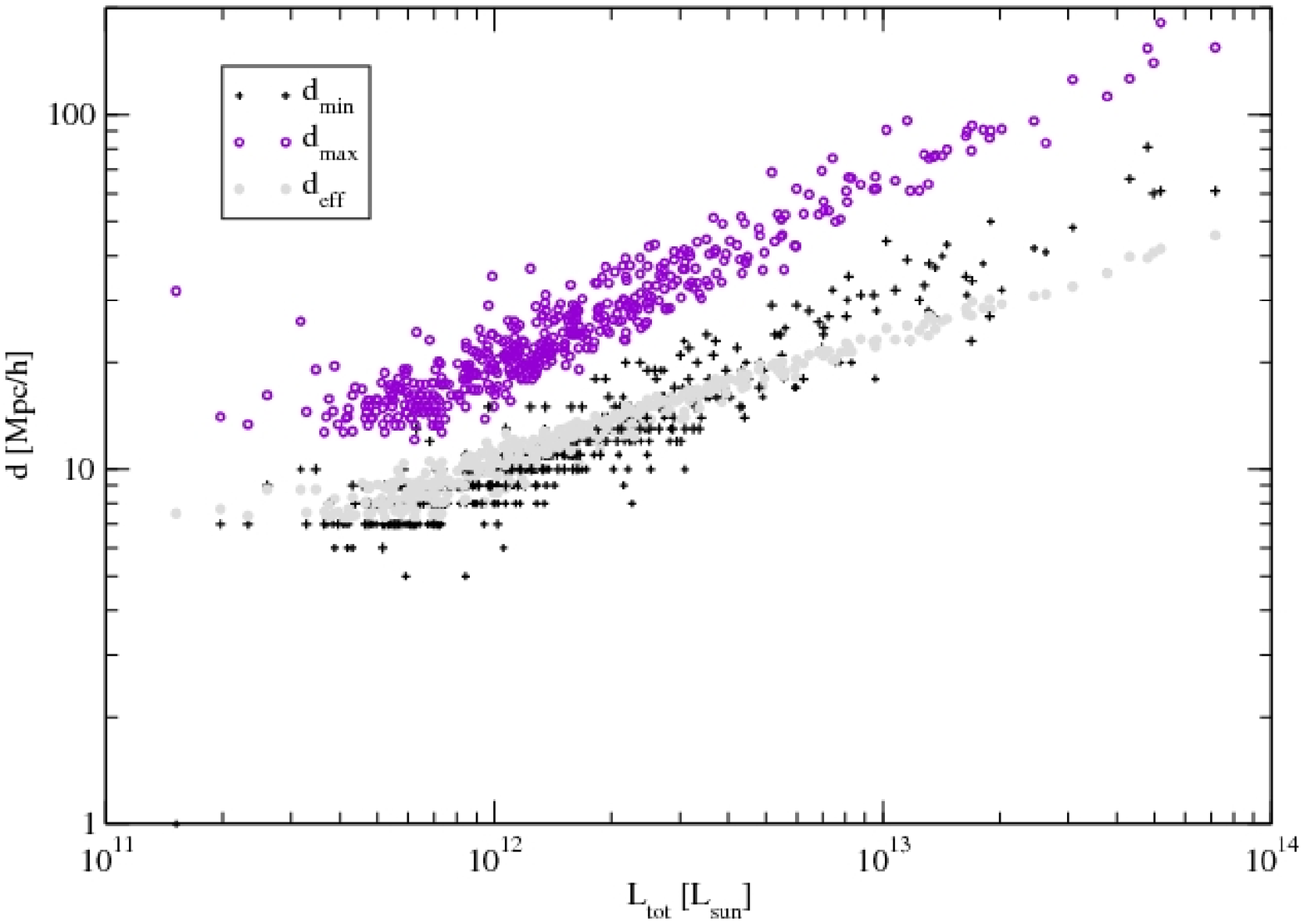}}
\resizebox{.48\textwidth}{!}{\includegraphics*{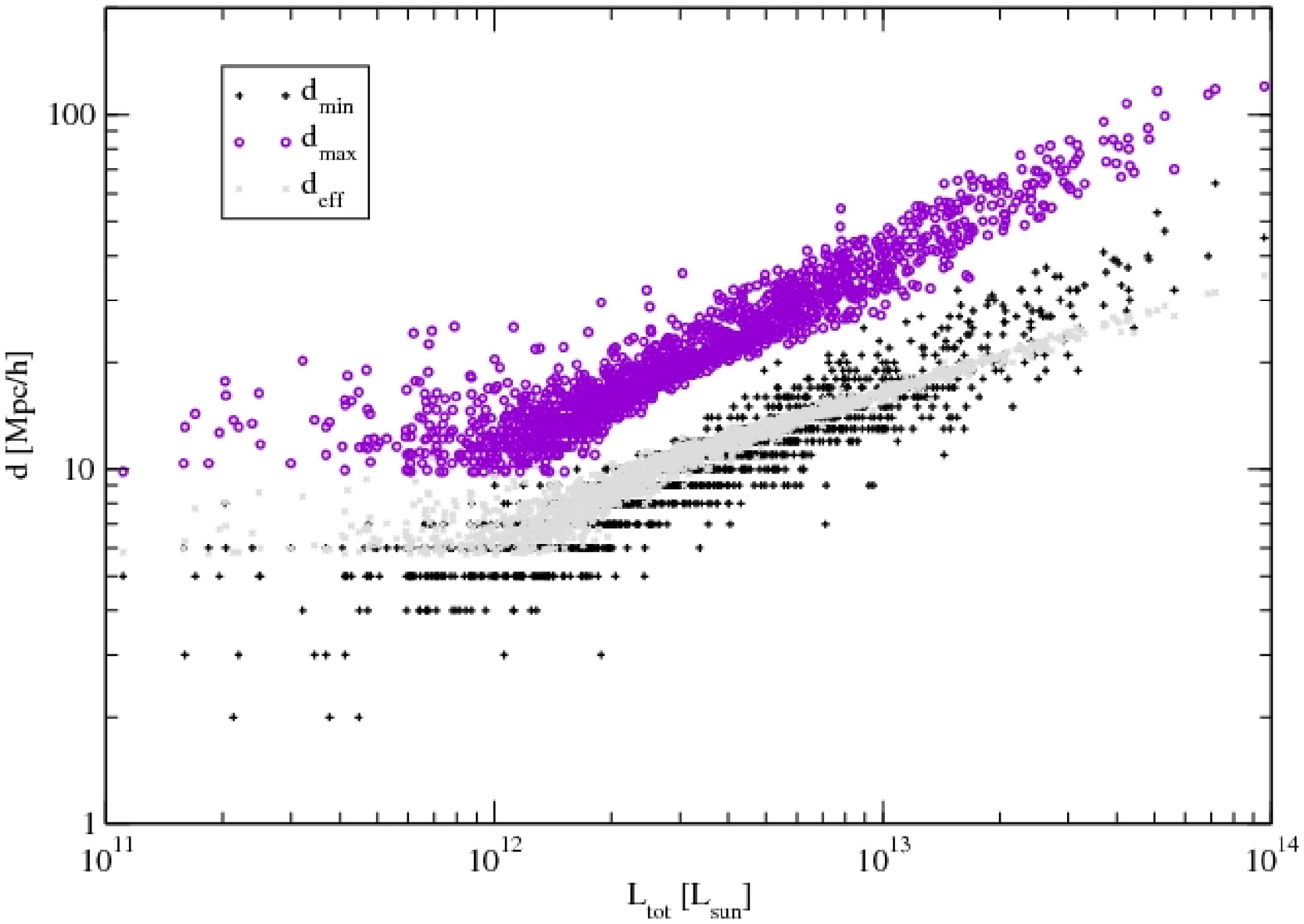}}
\\
\caption{The minimal, maximal and effective diameter of superclusters
  of various total luminosity.  The left panel shows data for 2dF
  superclusters, the right panel for Millennium Simulation
  superclusters.  }
\label{fig:9}
\end{figure*}

The general shape of the integrated luminosity function, found here
for the 2dFGRS and Millennium Simulation supercluster samples, is
rather close to the theoretical luminosity function calculated by
Oguri et al. (\cite{oguri04}) using the Press-Schechter
approximation.  Both real and simulated supercluster samples span an
interval of total luminosities exceeding two orders of magnitude,
similar to the range of luminosities discussed by Oguri et al. They
compared the model with the sample of superclusters of SDSS by E03a, which was 
compiled on the basis of the 2-dimensional density field. This sample covered
only one decade in total luminosity. The present paper gives improved 
possibilities for comparison. Also, the number of DF-clusters
found in the present analysis is close to the number of clusters in
simulations by Oguri et al.

\begin{figure*}[ht]
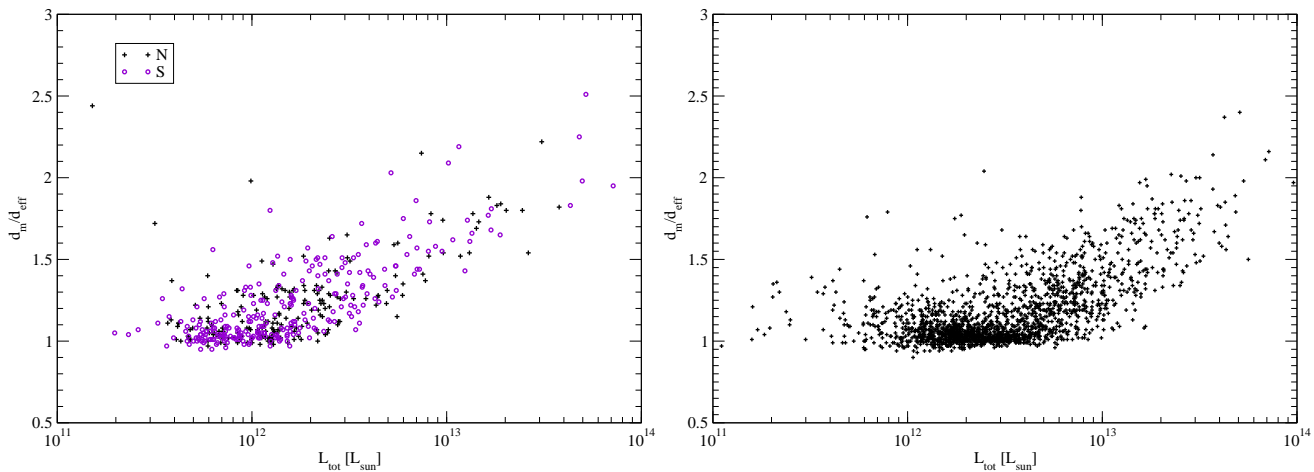

\centering
\resizebox{.48\textwidth}{!}{\includegraphics*{scl_NSgr8_rat-Ltot.eps}}
\resizebox{.48\textwidth}{!}{\includegraphics*{scl_M500F8_100_rat-Ltot.eps}}
\\
\caption{The ratio of the mean diameter to the effective diameter in
  superclusters of various total luminosity.  The left panel shows
  data for 2dF superclusters, the right panel for Millennium Simulation
  superclusters.  }
\label{fig:10}
\end{figure*}

\subsection{Sizes of superclusters}

Next we study the geometric properties of DF superclusters.  As an argument we
use the total luminosity of superclusters which was found by summing all
estimated total luminosities of member galaxies and groups/clusters in the
supercluster.  As geometric quantities we consider the minimal diameter
(along one of the coordinate axes, whatever is the smallest), the maximal
diameter of the supercluster defined as the length of the diagonal of the
rectangular box surrounding the supercluster along coordinate axes, and the
effective diameter -- the diameter of the sphere equal to the volume of the
supercluster (remember that the volume is defined as the number of cells which
have a density equal or greater than the threshold density, where the cell
size is 1~(\Mpc)$^3$).

The distribution of the maximal diameters of poor, medium, and rich
superclusters is shown in Fig.~\ref{fig:diam} for the 2dFGRS and Millennium
Simulation superclusters, median values and quantiles for superclusters of
different richness class are given in Table~\ref{tab:SCL-dat}.  As expected,
the maximal diameters depend strongly on the supercluster richness, both in
the real data and in the simulations.  Median values of maximal (and minimal)
diameters of superclusters of various richness are in good agreement with
semiaxis lengths for Abell superclusters (Jaaniste et al. \cite{ja98}) and
LCRS superclusters (Doroshkevich et al. \cite{dor01}) of similar richness.  An
even closer correlation exists between the effective diameter and the total
luminosity, shown in Fig.~\ref{fig:9}.  The scatter in this relationship is
the smallest and the relationship in log-log representation is linear.

Figure~\ref{fig:9} demonstrates one important aspect of our supercluster
catalogue: the total luminosities and sizes of superclusters have a rather
sharp lower limit.  This limit is determined by two selection parameters: the
density threshold and the volume threshold.  The density threshold excludes
systems of galaxies of low density from the sample, and the volume threshold
excludes small systems like single clusters.  We have used values as low as
possible in our supercluster selection in an attempt to include all galaxy
systems which could be classified as superclusters.  The analysis presented
above has shown that the choice of these selection parameters has been rather
successful: the sample of superclusters is fairly homogeneous.  With our
choice of selection parameters we have excluded about 60\% of galaxies.  As
noted above, these galaxies are located in less dense galaxy systems, and the
main galaxies of these poor systems have a much broader distribution of
luminosities and their mean luminosities are also considerably lower (see Paper
III for more details).  All this
confirms that we have selected a practically complete sample of galaxy systems
(within the observed region) which can be called superclusters.

\subsection{Symmetry and compactness of superclusters}

The axis ratios of superclusters have a mean value around unity in all three
directions.  More interesting is the dependence of the minimal, maximal and
effective diameters on the expected total luminosity of the supercluster,
shown in Fig.  \ref{fig:9}.  This Figure shows that all three diameters are
the larger the higher the total luminosity of the supercluster and the richer
the supercluster (see Table~\ref{tab:SCL-dat}).  The dependence is, however,
different for various diameters: the maximal diameter shows the fastest growth
with luminosity, and the effective diameter the slowest growth.  This
difference is due to a variation of the shape and compactness of superclusters
with differing luminosity.

The right panel of Fig.~\ref{fig:9} shows the same set of axis ratios
for superclusters found in the Millennium Simulation Mill.F8.  The
general trend is very close to the trend seen in the real
supercluster sample.  This similarity shows that geometric properties
of simulated superclusters fit real data very well.

\begin{figure*}[ht]
\centering
\resizebox{.48\textwidth}{!}{\includegraphics*{scl_NSgr8_Coff-Ltot.eps}}
\resizebox{.48\textwidth}{!}{\includegraphics*{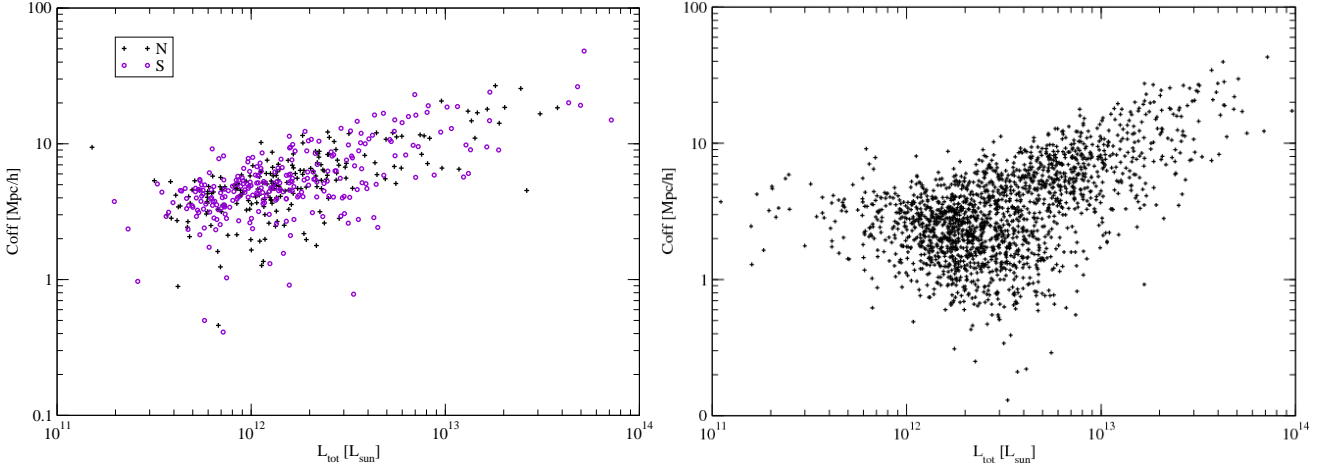}}
\\
\caption{The offset of the geometrical center from the
  dynamical one as a function of the total estimated luminosity of
  superclusters. Left panel shows 2dF superclusters, right panel
  Millennium Simulation superclusters. 
}
\label{fig:11}
\end{figure*}

The ratio of the mean diameter (the average of the extensions along
3-coordinate axes) 
to the effective diameter is plotted in Fig.~\ref{fig:10}, in the left
panel for 2dFGRS superclusters, and in the right panel for Millennium
Simulation superclusters.  This ratio depends on the compactness and
the filling factor of the system.  The ratio is the larger the more
empty space (more accurately, regions of space of luminosity density
below the threshold density) is located in the vicinity of the system
defined by galaxy filaments belonging to the system. The Figure shows
that in poor superclusters this ratio is close to unity (the minimum
by definition).  In other words, poor superclusters are rather
compact systems. Here a good example is the Local Supercluster, which
contains only one rich cluster, the Virgo cluster, close to its center,
and is surrounded by numerous galaxy filaments.  These filaments have
low luminosity density and fall outside the threshold density. Thus
systems like the Local supercluster are in our catalogue presented
only by their cores, which are compact and rather spherical, due to the
symmetrical smoothing of the density field.  The actual shape of
compact superclusters can be investigated using the distribution
galaxies or groups (see the next subsection).

The scatter of values about unity for poor superclusters is partly
due to errors of the mean diameter.  Supercluster sizes have been
found from density field cells with extreme values of coordinates
inside the threshold contour, they can only be determined with an accuracy
of $\pm 1$ \Mpc. The mean diameters of smallest systems are of the
order of 10 \Mpc, thus a relative error of about 10\% is expected for
the ratio of diameters of small systems.  In almost all rich
superclusters the mean diameter is much higher than the effective
diameter, i.e. these systems have a lower compactness and filling
factor.  This result agrees with that in E97, in which we showed that
rich superclusters have smaller fractal dimensions and are more
filament-like than poor superclusters.

Figure~\ref{fig:11} shows the offset of the geometrical center from the
dynamical one, again for the 2dFGRS and Millennium Simulation
superclusters. The center offset parameter was defined as follows:
$\Delta_o = ((x_0 - x_m)^2 + (y_0 - y_m)^2 + (z_0 - z_m)^2)^{1/2}$;
here $x_0, y_0, z_0$ are coordinates of the geometric center of the
supercluster, found on the basis of extreme values of coordinates of
the rectangular box containing the system, and $x_m, y_m, z_m$ are
coordinates of the dynamical center of the supercluster, defined by
the main cluster of the supercluster.  In the ideal case it would be
better to determine the dynamical center using the gravitational
potential field.  We have used instead the center of the most luminous
cluster near the highest density peak of the system.  Poor and medium
superclusters have relatively small offset of up to 10 \Mpc, but the
offset of very rich superclusters can reach values of up to 50 \Mpc.  In
other words, poor superclusters are fairly symmetrical, but luminous
superclusters are asymmetric: the system of filaments has different
lengths in various directions.  Actually filaments form a continuous
web, only in some parts the density of galaxies in filaments is lower
and falls below the threshold used to compile galaxy systems.  The
Figure also shows that there are no sharp boundaries between
superclusters of low and high luminosity -- i.e., the transition is smooth.

\subsection{The shape of superclusters}

To study the shape of superclusters we approximate the spatial
distribution of galaxies or groups in superclusters by a 3-dimensional
mass ellipsoid and determine its centre, volume and principal axes. 
In most cases our superclusters do
not form a regular body; however, these parameters can be used as a
first approximation to describe the density and alignments of the
elements of large-scale structure.

In the present study we use the classical mass ellipsoid
(see e.g. Korn \& Korn \cite{korn61}): 
\begin{equation}
\sum_{i,j=1}^3\left(\lambda_{ij}\right)^{-1}x_ix_j=5,  
\end{equation}
where
\begin{equation}
\lambda_{ij}={1\over{N_{gr}}}\sum_{l=1}^{N_{gr}} 
{(x^l_i-\xi_i)(x^l_j-\xi_j)},  
\end{equation}
is the inertia tensor for equally weighted groups, $N_{gr}$ is the 
total number of groups, and     $\xi_i={1\over
N_{gr}}\sum_{l=1}^{N_{gr}}x^l_i$ determines the Cartesian coordinates 
of the centre of mass of the system.

\begin{figure*}[ht]
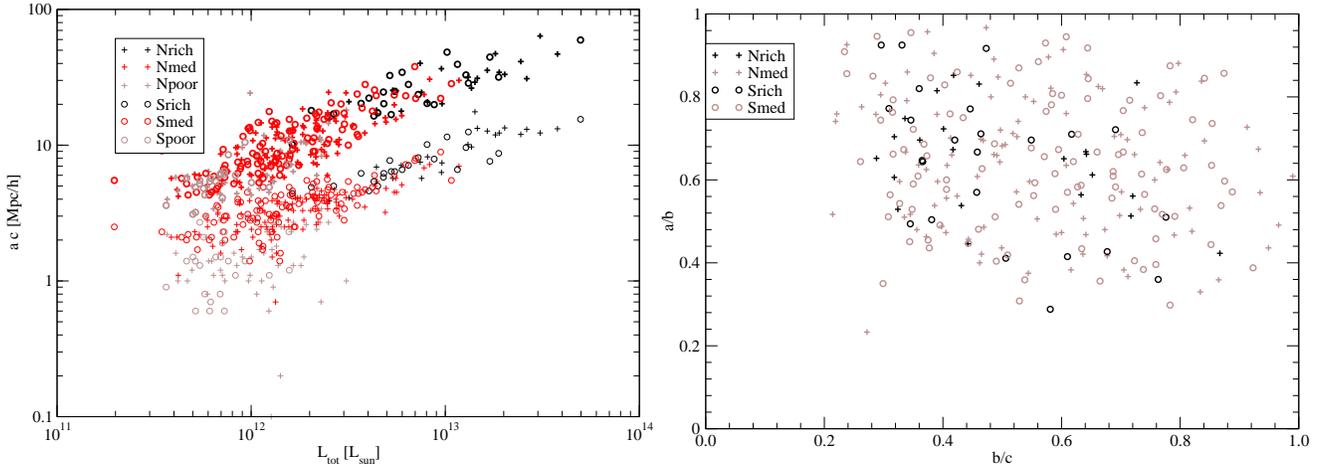

\centering
\resizebox{.48\textwidth}{!}{\includegraphics*{shape_NS_ac-Ltot.eps}}
\resizebox{.48\textwidth}{!}{\includegraphics*{shape_NS_ab-bc2.eps}}
\\
\caption{Left panel: the semi-minor and semi-major axes of the mass
  ellipsoid, $a$ and $c$, are plotted for superclusters of different
  total luminosities.  Bold symbols are for the semi-major axis, and thin
  ones for the semi-minor axis.  Superclusters of various richness are
  shown with different symbols. Right panel: axial ratios of
  superclusters  $a/b$ vs. $b/c$ for superclusters with at least 10
  member groups. Filamentary superclusters populate the upper left
  part of the figure, spherical superclusters the upper right part,
  and pancake-like structures the lower right section.  Triaxial
  superclusters are located in the central region. 
}
\label{fig:shape}
\end{figure*}

The formula determines a 3-dimensional ellipsoidal surface with the
distance from the centre of the ellipsoid equal to the rms deviation
of individual objects in the corresponding direction.  This method can
be applied for superclusters with $N_{gr} \ge 5$. The problems related
to the stability of the method and the influence of observational
errors have been discussed in Jaaniste et al. (\cite{ja98}).

To investigate the sensitivity of the method to the number of groups
we divided our sample of mass ellipsoids  into three richness
classes, Xrich with the number of groups $N_{gr} \ge 100$, Xmed with
$100 > N_{gr} \ge 10$, and Xpoor with $N_{gr} < 10$; here X denotes N
for the 2dFGRS Northern sample and S the Southern one. Superclusters
with the number of groups exceeding 10 are more-or-less uniformly
distributed in distance, whereas poor superclusters are all located at
distance exceeding 300~\Mpc\ (see also Fig.~\ref{fig:1}).  In
Figure~\ref{fig:shape} we show in the left panel the semi-minor axis $a$
and the semi-major axis $c$ as a function of the supercluster total
luminosity. We see that poor superclusters have a large scatter of
semi-axis length values. In particular, the length of the semi-minor
axis is for some superclusters less than 1~\Mpc. As all these
superclusters are distant objects with only a few groups/clusters in
the visibility window, we assume that this large scatter is due to
small number of objects.  Nearby superclusters of similar total
luminosity have considerably larger values of the semi-minor axis $a$ and
a much smaller scatter.  To avoid this uncertainty we
exclude samples of poor superclusters with $N_{gr} < 10$ from the further
analysis. 

Fig.~\ref{fig:shape} right panel shows the bivariate distribution of
axial ratios $a/b$ vs. $b/c$.  This figure can be compared directly with
Figures 9 - 11 of Wray et al. (\cite{wray06}) for samples of
simulated superclusters.  Wray et al. defined superclusters  as
clusters of rich  DM-clusters.  They used several linking
lengths to compile supercluster catalogues.  The largest linking
length yields percolating superclusters and cannot be compared with our
results, but the linking length $L = 5$~\Mpc\ corresponds approximately to
our accepted threshold density (Fig. 11 of Wray et al.).  The
comparison of our Figure~\ref{fig:shape} with the corresponding figure
by Wray et al. shows very good agreement.  Both distributions show that
there are no purely filamentary, spherical or pancake-like
superclusters: almost all superclusters are triaxial, with some
tendency toward filamentary character.

We further note that Figure~\ref{fig:shape} is akin to Figure~\ref{fig:9}.
Whereas the former shows the length of the semi-minor and semi-major axes $a$
and $c$ for superclusters of various total luminosities, the latter presents
minimal, maximal, and effective diameters of superclusters found from the
density field.  Semi-axes of the mass ellipsoid are about one-half to
one-third the size of their respective diameters, as would be expected.

This analysis complements our previous analysis based on the density field.
One small remark should be added: if the supercluster only barely meets our
threshold density criterion, and only its tip exceeds the threshold density
level, then, by definition, the density field above the threshold is almost
spherical in shape, due to the symmetrical smoothing law applied in the
determination of the density field. Thus the true shape of poor superclusters
cannot be determined from the density field alone.  Here the distribution of
objects within the supercluster can help, but only, if the number of objects
exceeds 10.

A similar conclusion on the shape of superclusters has been reached by
Jaaniste et al. (\cite{ja98}). Basilakos et al. (\cite{bpr01}) and
Basilakos (\cite{bas03}) applied shape-finders introduced by Sahni et
al. (\cite{sahni98}), to find the shape characteristics of PSCz and
SDSS superclusters.  PSCz superclusters were defined by the density
field method using rather large cell sizes of 5 and 10~\Mpc. SDSS
superclusters were found on the basis of Cut and Enhance clusters by
Goto et al. (\cite{goto02}).  In both cases this statistic suggests
that filaments dominate over pancakes.  Kolokotronis et
al. (\cite{kbp02}) compared shapes of Abell superclusters with
simulated superclusters using the same shape-finder statistics.  Again
a dominance of filamentary structures both in real and simulated
superclusters was found. Doroshkevich et al. (\cite{dor01}) determined
effective dimension for galaxy systems in high- and low-density regions, which
correspond to our supercluster and field samples.  For the supercluster sample
they found the dimension $1.6 \pm 0.2$, and for the field sample $1.0 \pm
0.17$, which correspond to sheets and filaments, respectively. Klypin et
al. (\cite{kees89}) investigated fractal dimensions of nearby superclusters
(Virgo, Coma) and galaxy filaments surrounding them.  For superclusters they
found effective dimensions 1.8 - 2.0, for inter-supercluster regions 1.3.

\begin{figure*}[ht]
\centering
\resizebox{.48\textwidth}{!}{\includegraphics*{scl_NSgr8_Dmean-Ltot.eps}}
\resizebox{.48\textwidth}{!}{\includegraphics*{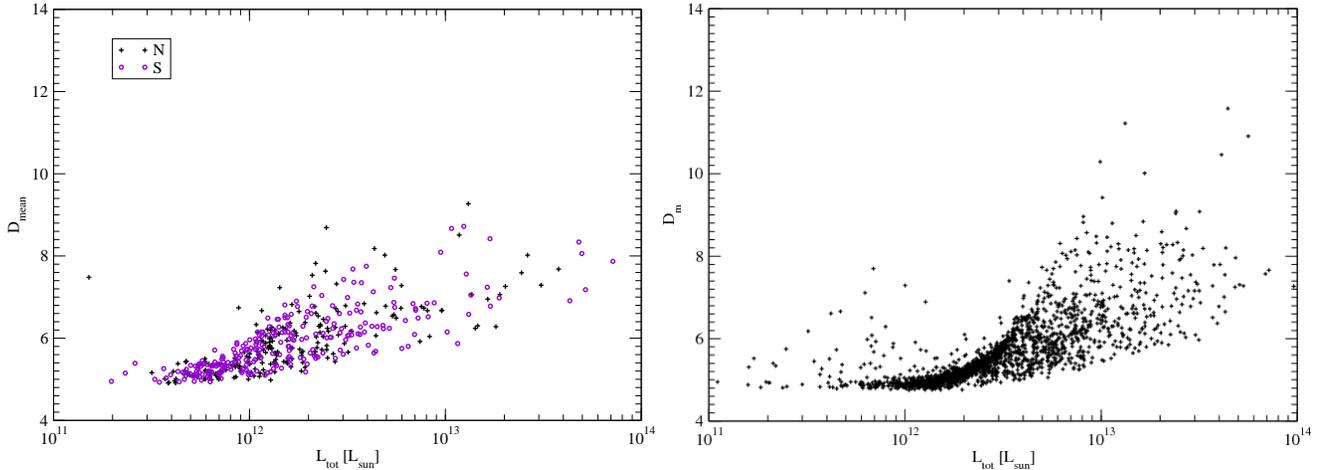}}
\\
\caption{Mean density of superclusters as functions of the total
 luminosity of the supercluster. Left panel shows data for 2dF
 samples, right panel for Millennium Simulation data, calculated with
 minimal volume 100~(\Mpc)$^3$.  }
\label{fig:13}
\end{figure*}

\begin{figure*}[ht]
\centering
\resizebox{.48\textwidth}{!}{\includegraphics*{scl_NSgr8_Dmax-Ltot2.eps}}
\resizebox{.48\textwidth}{!}{\includegraphics*{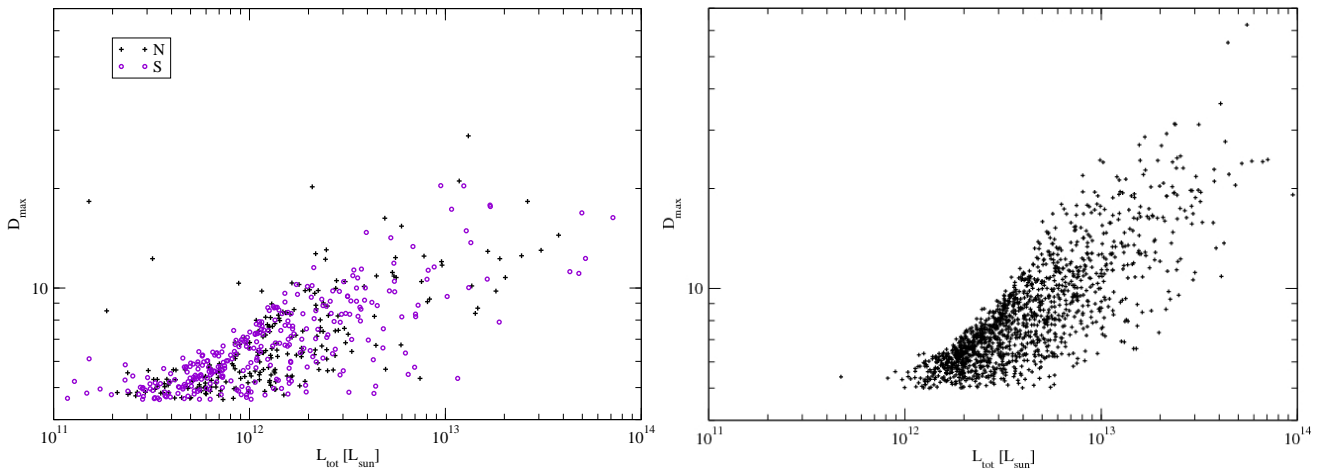}}
\\
\caption{Maximal density of superclusters as functions of the total
 luminosity of the supercluster. Left panel shows data for 2dFGRS
 samples, right panel for Millennium Simulation data, calculated with
 minimal volume 200~(\Mpc)$^3$. Notice the difference in the number of
 poor superclusters.}
\label{fig:14}
\end{figure*}

An inspection of the density fields of 2dFGRS samples shows that
actually the shape of rich superclusters is more complicated. It
cannot be presented by a simple ellipsoid.  The density field shows
that most superclusters are multi-branched, i.e. they consists of
numerous density knots arranged along several cluster chains.  This
morphology was detected already in the Perseus-Pisces supercluster by
J\~oeveer et al. (\cite{jet78}), Einasto et al. (\cite{ejs80}) and
Zeldovich et al. (\cite {zes82}).

\subsection{Densities of superclusters}

The mean luminosity density in superclusters is presented in Fig.~\ref{fig:13}
for 2dFGRS samples and for the Millennium Simulation sample.  The mean density
rises from 4.5 for poor to 6 -- 10 for rich superclusters.  We see also a
gradual increase of the mean density with increasing total luminosity.  This
fact is very important, since it demonstrates that rich superclusters are
dense systems, not just percolations of several loose systems.

The maximal luminosity density of superclusters (the smoothed density near the
dynamical center) is shown in Fig.~\ref{fig:14}.  Here the trend with
supercluster total luminosity is even more pronounced: very luminous
superclusters have also high-density peaks at their centers.  The maximal
density increases from 5 in poor superclusters to about 20 in rich ones. Note
that the maximal density of poor superclusters in many cases only marginally
exceeds the mean density (about 5), and both are very close to the threshold
density 4.6 used in the compilation of the supercluster catalogues.  This
demonstrates that poor superclusters are small density enhancements with a low
scatter of internal density.

Note, however, that the supercluster with the highest luminosity has not the
highest mean and maximal density, but a bit lower than the highest values.
This is the case both for the real as well as the simulated supercluster
sample.  This indicates that these largest superclusters are not the very
densest, but consist of a number of subunits of slightly lower mean and
maximal density.

\section{Discussion}

\subsection{Possible biases and errors of supercluster parameters}

There are two types of errors in our supercluster catalogue.
Systematic errors are due to uncertainties in the selection
algorithms and in the input parameters in the selection of
superclusters. Random errors are due to observational errors and errors
due to the small numbers of observed galaxies in superclusters.

Random errors in the galaxy positions are negligible.  More serious
are errors in redshifts and in magnitudes.  Large errors in redshift are
very seldom and move the galaxy completely outside the supercluster in
question, thus these errors influence the number of galaxies
in the system. Ordinary redshift errors increase the redshift scatter
of galaxies in groups and single galaxies inside superclusters.  The
scatter inside groups is taken into account during the group selection
procedure. In this paper we have used the mean redshifts of galaxies in
groups. The influence of random errors on the mean redshift of groups is
small, in most cases less than 1~\Mpc\ (if the redshift error is
transformed to distance error).  Redshift errors of single galaxies are
larger, but also do not exceed considerably  1~\Mpc, the size of an
elementary cell in the density field.  Since densities are smoothed
with an Epanechnikov kernel of radius 8~\Mpc, these errors do not
influence our results.

\begin{figure}[ht]
\centering
\resizebox{.48\textwidth}{!}{\includegraphics*{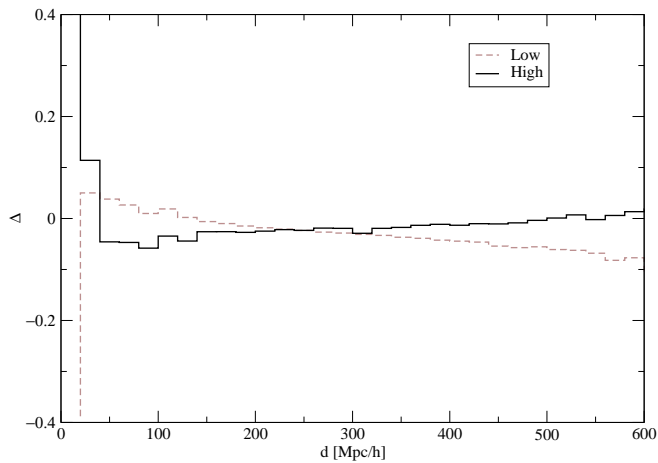}}
\caption{The mean difference of the density between density fields of
  Mill.A8 and Mill.F8, at various distances from the 'observer'. The
  solid line shows differences in overdensity regions, the dashed line
  in under-density regions. 
}
\label{fig:comp}
\end{figure}

Errors in magnitudes, in particular the corrections due to unobserved
galaxies, are the most serious random errors in our supercluster
catalogue. These errors increase the scatter of expected total
luminosities of superclusters.  More important are the errors of the
density field due to the use of corrected expected luminosities of
galaxies. As suggested by Basilakos et al. (\cite{bpr01}), the
smoothing of the density field may introduce a systematic error which
increases densities above the mean density and decreases densities
below the mean (see their Fig.~1). To investigate how serious this error is in
our case, we compared two density fields of the Millennium Simulation,
Mill.A8 and Mill.F8, and calculated the quantity
\begin{equation}
\Delta(r) = {\rho_F(r) - \rho_A(r) \over \rho_A(r)},
\end{equation}
where $\rho_A(r)$ is the mean density found for the original sample
Mill.A8, and $\rho_F(r)$ is the mean density found for the simulated
2dF sample. Mean values were found for a series of distance intervals
from the observer.  The results of the calculation are shown in
Fig.~\ref{fig:comp}. The trend is the same as found by Basilakos et
al., but the errors are about 10 times smaller. This difference in errors
is probably due to the use of very different cell sizes: in our case
the size was 1~\Mpc, whereas Basilakos et al. used cell sizes 5 and
10~\Mpc. In other words, deriving the density field using a small cell size
and calculating from flux-limited galaxy data does not introduce
noticeable systematic errors.  Random errors in the density field are
larger, as seen from the comparison of supercluster total luminosities
found for samples Mill.A8 and Mill.F8 (see Paper I).

The most serious errors in our catalogue are due to the small number of
galaxies observed in distant superclusters.  As seen from Figure~\ref{fig:1},
the number of galaxies observed in superclusters has a lower limit which
decreases with increasing distance according to the same law as found by
shifting groups of galaxies to larger distance (see T06).  At the outer limit
of our galaxy sample at $z = 0.2$ the lower limit of the number of galaxies
observed in superclusters approaches $N_{gal} = 1$.  It is clear that one
galaxy is not sufficient to define a supercluster, even a poor one. Thus we
have excluded superclusters with a number of groups below 3 from our list.
Furthermore, the catalogue was divided into two parts, the main catalogue A
and the supplementary catalogue B. The main catalogue has systems up to
distance 520~\Mpc, and the supplementary one even more distant objects.  We
have found the luminosity function of superclusters separately for the main
and the supplementary supercluster sample (see Fig.~\ref{fig:lumf}).  The
luminosity function of the supplementary sample has a larger scatter and lies
higher for rich superclusters.

{
\scriptsize
\begin{table}[ht] 
\caption{Identification of rich 2dF Northern superclusters 
in another supercluster catalogues} 
\begin{tabular}{rrrrrrrrl} 
\hline 
\\
ID & ID$_W$ &  ID$_{ACO}$ & $ N_{cl}$ & $N_{gr}$ & $N_{ACO}$ & 
$N_{2dF}$ & $N_X$ & N \\
1  &  2     &    3        &    4      &    5     &    6      &   7 & 8 & 9 \\
\hline 
\\
   13 &     07 &     82 & 10 & 1145 & 4 &  11 & & \\
   20 &     01 &     88 &  2 &  556 & 1 & 7   & 2 & 1. \\
   37 &        &    265 &  9 &  359 & 1 & 2   & & \\
   76 &  07,08 &    100 &  5 &  420 & 1 & 4   & & \\
   77 &        &     91 &  2 &  315 & 1 & 3   & 1 & 2.  \\
   78 &        &        &  5 &   57 & 1 & 2   & & \\
   82 &  08,11 &        &  3 &  187 & 1 & 10  & & \\
   92 &     08 &100,265 &  3 &  315 & 2,1 &  3,4  & & 3. \\
   97 &     08 &    265 &  5 &  129 & 2 & 5   & & \\
   99 &        &        & 13 &  472 & 8 & 14  & & \\
  108 &        &        & 24 &  169 & 1 & 2   & & \\
  120 &        &        & 19 &  207 & 1 & 1   & & \\
  136 &     14 &        &  2 &  251 & 1 & 2   & & \\
  152 &     06 &111,126 & 18 & 3591 & 2,7 & 2,40   & 5 & 4. \\
  170 &     10 &        &  8 &  415 &   &     & & \\
  205 &     12 &        &  5 &  215 & 1 & 2   & & \\
  220 &        &        & 16 &  426 & 2 & 2   & & \\
\hline
\label{tab:SCL-ID1}
\end{tabular}
Colums:\\
1 - 2dFGRS supercluster ID number,\\
2 - ID number by Erdogdu et al.,\\
3 - Abell supercluster ID number (E01),\\
4 - number of DF clusters,\\
5 - number of 2dFGRS groups + field galaxies, \\
6 - number of Abell clusters, \\ 
7 - number of 2dFGRS groups (T06),\\
8 - number of X-ray clusters,\\
9 - name (see Notes).\\
Notes: 1. Sextans; 2. Leo-Sextans; 3. Leo-A; 4.Virgo-Coma \\
\end{table}
}

{
\scriptsize
\begin{table}[ht] 
\caption{Identification of rich 2dF Southern superclusters
in another supercluster catalogues} 
\begin{tabular}{rrrrrrrrl} 
\hline 
\\
ID & ID$_W$ &  ID$_{ACO}$ & $N_{cl} $ & $N_{gr}$ & $N_{ACO}$ &
 $N_{2dF}$ & $N_X$ & N \\
1  &  2     &    3        &    4      &    5     &    6      &   7 & 8 & 9 \\
\hline 
\\
    5 &     04 &     10 &  5 &  952 & 1 & 5 & 5 & 1.  \\
   10 &        &        & 17 &  535 & 2 & 5 & & \\
   11 &        &        &  3 &  101 & 1 & 1 & & \\
   19 &        &        &  2 &   91 & 1 & 5 & & \\
   34 &     16 &    5,9 & 24 & 3175 & 2,8 &  9,26 &  6 & 2. \\
   51 &     18 &        &  7 &  272 & 1 &  3 &  & \\
   60 &        &        &  4 &  132 & 1 &  1 &  & \\
   78 &        &        & 20 &  254 & 1 & 1  & & \\
   84 &        &        &  3 &  225 & 3 & 4  & & \\
   87 &        &        &  4 &  166 & 4 & 8  & & \\
   88 &        &        &  2 &  105 & 1 & 1  & & \\
   94 &     18 &        & 15 &  245 & 1 & 2  & & \\
  109 &     15 &    232 &  2 &  249 & 2 & 7  & & \\
  115 &        &        & 10 &  265 & 1 & 2  & & \\
  116 &        &    232 &  3 &  230 & 2 & 4  & & \\
  126 &     15 &        &  3 &  291 & 2 & 8  & & \\
  152 &        &        &  4 &  180 & 1 & 1  & & \\
  148 &        &        & 11 &  328 & 4 & 9  & & \\
  153 &     15 &        &  8 &   64 & 1 & 5  & & \\
  167 &     06 &     49 &  2 &  771 & 2 & 2  & 1 & \\
  190 &        &        &  5 &  122 & 1 & 4  & & \\
  200 &        &        &  6 &  155 & 1 & 1  & & \\
  204 &        &    190 &  5 &  342 & 2 & 5  & & \\
  217 &        &        & 42 &  938 & 4 & 12 & & \\
  222 &        &    199 &  2 &  473 & 1 &  4 & & \\
  240 &        &        &  6 &  171 & 3 &  6 & & \\
  267 &        &        &  6 &  173 & 2 &  5 & & \\
  276 &        &        &  8 &  371 & 1 &  2 & & \\
  303 &        &        &  5 &   71 & 1 &  2 & & \\
\hline
\label{tab:SCL-ID2}
\end{tabular}
Notes: 1. Pisces-Cetus, 2. Sculptor\\ 
\end{table}
}

\subsection{Comparison with other supercluster catalogues} 

In Tables~\ref{tab:SCL-ID1} and \ref{tab:SCL-ID2} we give the identification
of superclusters from our lists with the list by Erdogdu et al.
(\cite{erd04}) obtained using the Wiener Filter method, and with Abell
superclusters by Einasto et al. (\cite{e2001}).  For superclusters which have
no partners in the Abell supercluster list, we give the number of Abell
clusters within the volume of the supercluster.  The reason for the absence of
the supercluster in the Abell supercluster list can be either a too low number
of Abell clusters in the supercluster volume (identified in Abell supercluster
search as isolated Abell cluster), or the clusters was too distant to be
included in the Abell supercluster catalogue. The commonly used name given in
the ninth columns is according to Abell supercluster catalogue by E01.

This comparison shows that there exist no one-to-one relationship between
superclusters of our lists and the lists by other authors.  Differences are due to
the usage of different techniques in supercluster definition and to the usage of 
different observational data.

Erdogdu et al. used a much larger and variable cell size when deriving the
smoothed density field, thus their method has lower resolution than ours.
Taking this into account, it is not surprising that several superclusters found
with the Wiener filtering method are split into separate systems in our list.
For instance, the Wiener Northern supercluster 08 combines our superclusters
76, 82, 92 and 97, and the Wiener Southern superclusters 15 and 18 are
combinations of our superclusters 109, 126 and 51, 94, respectively. If we use
a lower threshold density, then at a certain level our method also joins these
superclusters to single systems.  We repeat that the actual web of
superclusters and filaments is a continuous one, and it is matter of
convention how to join parts of this web to particular systems.

Abell superclusters were found using only Abell clusters as tracers -- no
individual galaxy 
information was used.  Also the linking length used to combine clusters into
superclusters corresponds to a much lower threshold density in our method. Due
to these differences in data and method, one would expect to see more
differences in the results.  Actually there exists a rather good
correspondence between our and Abell superclusters. Most notable differences
are that Abell supercluster 265 is divided in our list into three entries (37, 92
and 97) in the Northern region, and that Abell 232 is a combination of our 109 and
116 in the Southern region. But there are also examples in another
direction: the most prominent supercluster in the Northern region, 152 in our
list and 06 in the Wiener supercluster list, is partly divided into Abell
superclusters 111 and 126.

\subsection{Comparison of real and simulated superclusters}

The comparison of properties of simulated superclusters with real
superclusters shows that in most cases simulated superclusters have relations
between various parameters and the total luminosity which are almost
identical to similar relations for real superclusters.  There are some
differences: the luminosity and the multiplicity functions of the Millennium
Simulation superclusters do not contain as many very rich superclusters as is 
found in the 2dFGRS sample.  We shall discuss these differences in more detail
elsewhere (Einasto et al. \cite{esh06}, Saar et al. \cite{see06}).

The similarity of results for real and simulated superclusters has
several consequences. First of all, it shows that our procedures to
define superclusters and their parameters are rather robust and yield
stable results.  Secondly, it shows that simulations have reached a
stage which produces superclusters of galaxies with properties very 
close to the observations.

\begin{figure*}[ht]
\centering
\resizebox{.48\textwidth}{!}{\includegraphics*{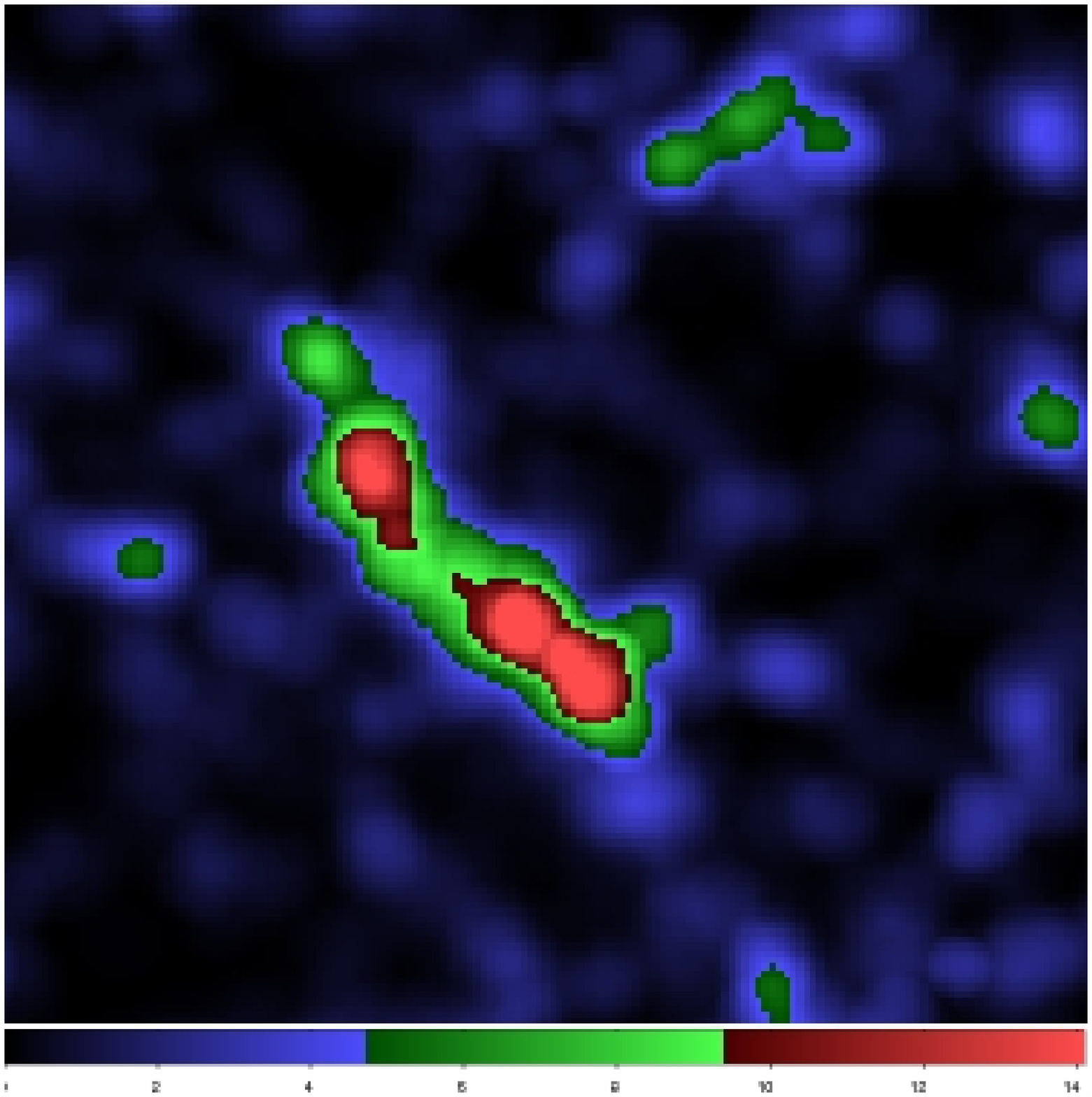}}
\resizebox{.48\textwidth}{!}{\includegraphics*{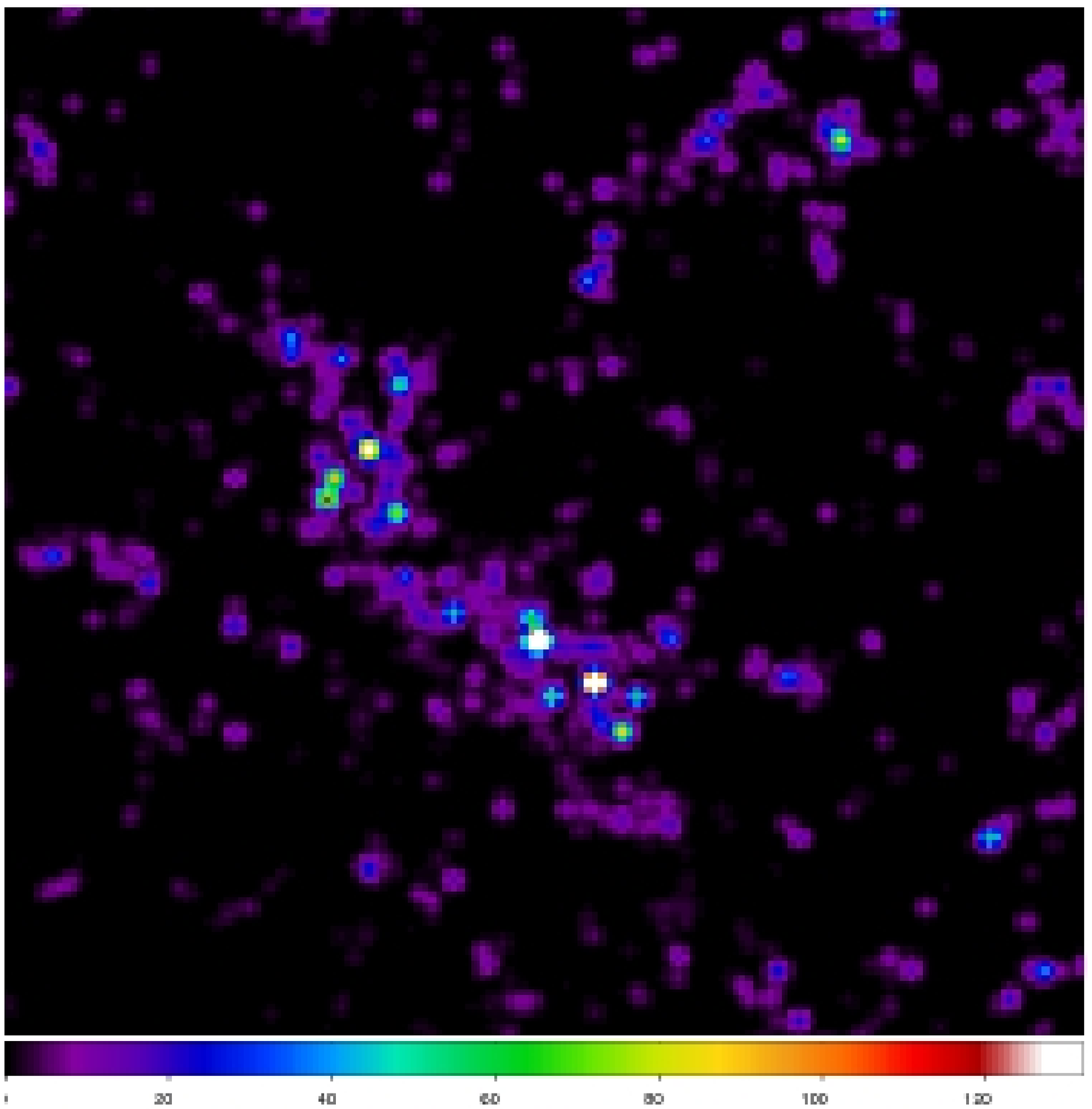}}
\\
\caption{The density field through the supercluster SCL126.  Left
  panel shows the low-resolution field, the right panel the
  high-resolution field.  In the low-resolution field the contour
  limiting green and blue regions corresponds to threshold density 4.6
  in units of the mean density, which separates supercluster regions
  from low-density ones.  
}
\label{fig:scl}
\end{figure*}

\subsection{Superclusters and the cosmic web}

Superclusters are large-scale density enhancements in the cosmic web,
the supercluster-void network.  Rich and very rich superclusters
contain rich clusters of galaxies in a relatively small volume, thus
such objects are easily detected.  Their properties depend on the
method of selection of superclusters, but the principal properties are
fairly stable. As an example we show in Figure~\ref{fig:scl} the low-
and high-resolution density fields through the richest supercluster in
the 2dFGRS Northern region, SCL126. This supercluster has been
called the Sloan Great Wall (Vogeley et al. \cite{vogeley04}, Gott et
al. \cite{gott05}, Nichol et al. \cite{nichol06}). Actually, the
wall-like shape is due to the enhancement of the structure, as at this
distance the observed spatial density of galaxies in the flux limited 2dFGRS
and SDSS samples has a maximum.  Praton, Melott, \& McKee (\cite{praton97})
refer to this as the "Bull's Eye" effect.  We see that the density contour 4.6
includes only the main body of the supercluster, the outlying parts and
galaxy filaments (seen in the high-resolution density field) remaining 
outside the supercluster volume.  Similarly, in poor superclusters, the
supercluster volume contains only the central core of the
supercluster.

Very rich superclusters have been found also in our vicinity outside
the 2dFGRS and SDSS surveys: the Shapley Supercluster and the
Horologium-Reticulum Supercluster (see Proust et al. \cite{proust06},
Fleenor et al. \cite{fleenor05} and Ragone et al. \cite{ragone06} and
references therein).

Most superclusters discussed in this paper are poor. They form small and
medium density enhancements in the cosmic web.  The web is a continuous
network of filaments and sheets of galaxies, and there exist density
enhancements of very different scale and luminosity. This continuous network
of filaments was investigated recently by Sousbie et al. (\cite{sousbie06}).
They found an algorithm which allows one to find the 3-dimensional filamentary
skeleton of the equatorial slice of the SDSS.  This slice is overlapped by the
2dFGRS Northern region and is centered around the supercluster SCL126, the
Sloan Great Wall. The density field method with a certain threshold density
allows one to identify as superclusters only their denser parts.  As shown by
Sousbie et al. and seen in Figure~\ref{fig:scl}, the filamentary network
continues outside superclusters.

There exist numerous investigations concerning the shape of structural
elements of the cosmic web: Bond et al. (\cite{bkp96}), Doroshkevich
et al. (\cite{dor01}, \cite{dor04}), Kasun \& Evrard \cite{kasun04}, Shen et
al. (\cite{shen05}), Sousbie et al. (\cite{sousbie06}), to name only a
few studies.  Doroshkevich et al. (\cite{dor01})  find that about 40 - 50 \% of
all galaxies belong to the ``wall'' (i.e. supercluster) population.  This is
in very good agreement with our results: we find that for threshold density
4.6 the fraction of galaxies in the supercluster population is 42 \% (in both
2dFGRS regions).   The volume filled with superclusters is much lower: 3.2 \%
in the Northern region, and 3.5 \% in the Southern one. This volume filling
factor is two times lower than found by E05b for simulated supercluster
population.  The difference my be explained by the use of lower threshold
density by E05b (2.66 in units of the mean density).

\section{Conclusions}

We made an analysis of properties of superclusters listed in Paper I.
Properties of real superclusters have been compared with properties of
simulated superclusters from the Millennium Simulation, using identical
procedures to collect galaxies to superclusters.

\begin{itemize}

\item{} We find that our sample of superclusters forms a homogeneous sample of
  galaxy systems, where properties of superclusters smoothly change with the
  total luminosity and multiplicity of the supercluster.  Using the
  multiplicity we divide superclusters into four richness classes: poor,
  medium, rich and extremely rich.
  
\item{} We investigated the shape of superclusters using groups of galaxies
  located inside superclusters, and the configuration of the density field
  above threshold used to define superclusters.  We find that rich
  superclusters are more asymmetrical and have a smaller filling factor than
  poor ones. The asymmetry is characterized by the offset of the geometrical
  mean center from the dynamical one, defined as the center of the main
  cluster. Another manifestation of the asymmetry is the ratio of the mean
  diameter to the effective diameter (the diameter of a sphere equal to the
  volume of the supercluster); this ratio characterizes the filling factor and
  the degree of filamentarity of the system.

\item{} We find that the mean and the maximal densities of
  superclusters increase when going  from poor to rich
  superclusters. This fact demonstrates that rich superclusters are
  not due to artificial percolation of poorer superclusters: they form
  a class of galaxy systems with properties continuously changing with
  supercluster richness.
  
\item{} We calculated the luminosity and the multiplicity functions of
  superclusters; both span over two decades in luminosity and spatial density.
  The richest superclusters of the 2dFGRS sample contains up to 60
  DF-clusters, whereas the the richest superclusters of simulated
  superclusters contains only up to 20.  This is the main difference of
  simulated supercluster sample from observations.

\item{} The comparison of properties of 2dFGRS superclusters with 
  those of superclusters found for the Millennium Simulation
  shows that almost all geometric properties of simulated
  superclusters are very close to similar properties of real
  superclusters of the 2dFGRS sample.

\end{itemize}

\begin{acknowledgements}
  
  We are pleased to thank the 2dFGRS Team for the publicly available final
  data release. The Millennium Simulation used in this paper was carried out
  by the Virgo Supercomputing Consortium at the Computing Centre of the
  Max-Planck Society in Garching. This research has made use of SAOImage DS9,
  developed by Smithsonian Astrophysical Observatory. The semi-analytic galaxy
  catalogue is publicly available at
  http://www.mpa-garching.mpg.de/galform/agnpaper.  The present study was
  supported by Estonian Science Foundation grants No.  4695, 5347 and 6104 and
  6106, and Estonian Ministry for Education and Science support by grant TO
  0060058S98. This work has also been supported by the University of Valencia
  through a visiting professorship for Enn Saar and by the Spanish MCyT
  project AYA2003-08739-C02-01.  D.T. was supported by the US Department of
  Energy under contract No.  DE-AC02-76CH03000.  J.E.  thanks
  Astrophysikalisches Institut Potsdam (using DFG-grant 436 EST 17/2/05) for
  hospitality where part of this study was performed.  In this paper we make
  use of R, a language for data analysis and graphics (Ihaka \& Gentleman
  \cite{ig96}).

\end{acknowledgements}

\end{document}